\documentclass[preprint, showpacs,preprintnumbers,amsmath,amssymb,nofootinbib]{revtex4}
\usepackage{amssymb}
\usepackage{amsfonts}
\usepackage{CJK}

 \usepackage{epsf}
 \usepackage{graphicx}
\usepackage{float}
\begin{document}
\newcommand{\be}[1]{\begin{equation}\label{#1}}
 \newcommand{\ee}{\end{equation}}
 \newcommand{\bea}{\begin{eqnarray}}
 \newcommand{\eea}{\end{eqnarray}}
 \newcommand{\bed}{\begin{displaymath}}
 \newcommand{\eed}{\end{displaymath}}
 \def\disp{\displaystyle}

 \def\gsim{ \lower .75ex \hbox{$\sim$} \llap{\raise .27ex \hbox{$>$}} }
 \def\lsim{ \lower .75ex \hbox{$\sim$} \llap{\raise .27ex \hbox{$<$}} }

\title{Probing modified gravity theories with the Sandage-Loeb test}

\author{Zhengxiang Li$^{1}$, Kai Liao$^{1}$, Puxun Wu$^{3}$, Hongwei Yu$^{2, 3,} \footnote{Corresponding author: hwyu@hunnu.edu.cn}$, and Zong-Hong Zhu$^{1}$}

\address{$^1$Department of Astronomy, Beijing Normal University,
Beijing 100875, China
\\$^2$Department of Physics and Key
Laboratory of Low Dimensional Quantum Structures and Quantum Control
of Ministry of Education,
\\Hunan Normal University, Changsha, Hunan 410081, China
\\$^3$Center of Nonlinear Science and Department of Physics, Ningbo
University,  Ningbo, Zhejiang 315211, China }

\begin{abstract}
In this paper, we quantify the ability of a future measurement of
the Sandage-Loeb test signal from the Cosmic-Dynamic-Experiment-like
spectrograph to constrain some popular modified gravity theories
including the Dvali-Gabadadze-Porrati braneworld scenario, $f(R)$
modified gravity and $f(T)$ gravity theory. We find that the
Sandage-Loeb test measurements are able to break degeneracies
between model parameters markedly and thus greatly improve
cosmological constraints for all concerned modified gravity theories
when combined with  the latest observations of CMB--shift parameter.
However, they yield almost the same degeneracy directions between
model parameters as that from the distance ratio data derived from
the latest observations of the cosmic microwave background and
baryonic acoustic oscillations. Moreover, for the $f(R)$ modified
gravity, the Sandage-Loeb test could provide completely different
bounded regions in model parameters space as compared to CMB/BAO and
thus supplement strong complementary constraints.

\end{abstract}


 \maketitle
 \renewcommand{\baselinestretch}{1.5}

\section{INTRODUCTION}
In the past decade or so, an accelerating expanding universe was first
indicated by observations of type Ia supernova (SNe
Ia)~\cite{Riess0,Perlmutter} and subsequently confirmed by several
other surveys such as the cosmic microwave background (CMB)~\cite{CMB},
and baryonic acoustic oscillation (BAO)~\cite{BAO}. This ostensibly
counterintuitive behavior of the universe is usually attributed to a
presently unknown component, called dark energy, which exhibits
negative pressure and dominates over the matter-energy content of
our universe. So far, the simplest candidate for dark energy is the
cosmological constant and the standard cosmological model based on
it (dubbed $\Lambda$CDM) is in accordance with almost all the
existing cosmological observations. However, dark energy is not the
only explanation for the current cosmic acceleration, modified
gravity theories in which gravitational interaction deviates from
Einstein's theory of gravity can also account for this apparently
unusual phenomenon. So far, both models derived from introducing an
exotic component like dark energy and those established by modifying
Einstein's theory of gravity can survive the above-mentioned
observations. If one wants to place more comprehensive cosmological
constraints on a possible model or distinguish between dark energy
and modified gravity theories, it is crucial to measure the
expansion rate of universe at many different redshifts. Among the
known probes, the CMB probes the rate of the expansion at redshift
$z\sim1100$, while for much lower redshift ($z\leqslant2$) we could
rely on weak lensing, BAOs, and the most noticeably, luminosity distance
measurements of SNe Ia and some other probes. In particular, a new
cosmological window would open if we could measure the cosmic
expansion directly within the ``redshift desert", roughly
corresponding to redshifts $2\leqslant z\leqslant5$.

Currently, the detailed dynamics of the accelerated expansion is
still not well known. However, one could expect the redshift of any
given object to exhibit a specific time evolution in an underlying
cosmological model. The observation of this evolution performed over
a given time interval could not only be a direct probe of the
dynamics of the expansion, but also has the advantage of not
depending on a determination of the absolute luminosity of the
observed source. Allan Sandage first proposed the possible
application of this kind of observation as a cosmological
tool~\cite{Sandage}. However, only measurements performed at time
interval separated by more than $10^7$ years could have detected the
cosmic signal with the technology at available that time. Over the
past decades, the importance of this method was stressed
again~\cite{Rudiger,Lake}. Later, Loeb revisited these ideas and
argued that spectroscopic techniques developed for detecting the
reflex motion of stars induced by unseen orbiting planets could be
used to detect the redshift variation of quasar stellar object (QSO)
Lyman-$\alpha$ absorption line~\cite{Loeb}. He also concluded that
it is conceivable that the cosmological redshift variation in the
spectra of some suitable source could be detected in a few decades.
Therefore, this method is usually referred as the ``Sandage-Loeb"
(SL) test. More recently, by using the Green Bank Telescope (GBT)
observation over 13.5 years, precise measurements for the secular
redshift drift of 10 H{\scriptsize I} 21 cm absorption line systems
spanning $z=0.09-0.69$ were obtained~\cite{Darling}. These surveys
were announced as direct measurements of the cosmic acceleration and
an error-weighted mean secular redshift drift of
$<\dot{z}>=(-2.3\pm0.8)\times10^{-8}~\mathrm{yr}^{-1}$,
corresponding to an acceleration of
$<\dot{v}>=-5.5\pm2.2~\mathrm{m}\cdot \mathrm{s}^{-1}\cdot
\mathrm{yr}^{-1}$, was achieved. Encouragingly, the cosmic
acceleration could be directly measured in $\sim125$ years with
current telescopes or in $\sim5$ years using a Square Kilometer
Array~$\footnote{http://www.skatelescope.org/}$.

An investigation of the expected cosmological constraints
from the SL test for a constant dark energy equation of state was
performed by Corasaniti {\it et al.}~\cite{Corasaniti}. Later, several extended analysis for some other
popular competing models including Chaplygin gas, holographic dark
energy and the new agegraphic and Ricci dark
energy~\cite{Balbi,Hongbao,Jinfei}, were accomplished. More
recently, the ability of a future measurement of the SL signal from
a Cosmic-Dynamics-Experiment-like (CODEX)~\cite{CODEX} spectrograph
to constrain a dynamical dark energy (CPL
parametrization~\cite{CPL}) was quantified by Martinelli {\it et
al.}~\cite{Martinelli}. Alongside the full CMB mock data set with
noise properties consistent with Plank-like~\cite{Plank} experiment,
they demonstrated that the SL test measurements could be able to
break degeneracies between expansion parameters and improve
cosmological constraints greatly.

In this paper,  we perform, along the line of
Ref.~\cite{Martinelli}, a joint analysis by taking the latest
observations of CMB--shift parameter into account to investigate the
constraining power of the SL test on model parameters of modified
gravity theories including the Dvali-Gabadadze-Porrati (DGP)
brane-world scenario~\cite{DGP},  the $f(R$)
 gravity (see Refs.~\cite{FR1,FR2} for recent reviews) and the $f(T$)
gravity~\cite{FT}. Moreover, analysis with the distance ratios
derived from the latest CMB and BAO observations (labeled as
CMB/BAO), which are deemed to be more suitable than the primitive
CMB data for examining non-standard dark energy models, are also
taken into consideration for comparison.

\section{DATA SETS}
In this section, we give brief descriptions for the data sets.
\subsection{Sandage-Loeb Test}
We begin with reviewing the basic theory necessary to derive the
expected redshift drift over a  time interval in a given
cosmological model. In a Friedmann-Lema\^{\i}tre-Robertson-Walker
(FLRW) expanding universe, the radiation emitted by a source which
does not possess any peculiar motion at time $t_s$ and observed at
time $t_0$ experiences a redshift $z_s$ which is connected to the
expansion rate through the scale factor $a(t)$ as
\begin{equation}\label{eq1}
1+z_s(t_0)=\frac{a(t_0)}{a(t_s)}.
\end{equation}
After a time interval $\Delta t_0$ (corresponding to $\Delta t_s$
for the source), it becomes
\begin{equation}\label{eq2}
1+z_s(t_0+\Delta t_0)=\frac{a(t_0+\Delta t_0)}{a(t_s+\Delta t_s)}.
\end{equation}
With an adequate time interval between observations, we can measure
the observed redshift variation
\begin{equation}\label{eq3}
\Delta z_s=\frac{a(t_0+\Delta t_0)}{a(t_s+\Delta
t_s)}-\frac{a(t_0)}{a(t_s)}.
\end{equation}
By keeping the first order in $\Delta t/t$, this difference can
be re-written as
\begin{equation}\label{eq4}
\Delta z_s\simeq \Delta
t_0\bigg[\frac{\dot{a}(t_0)-\dot{a}(t_s)}{a(t_s)}\bigg].
\end{equation}
Conveniently, this redshift variation is usually expressed in terms
of a spectroscopic velocity shift, i.e.,
\begin{equation}\label{eq5}
\Delta v=\frac{c\Delta z_s}{1+z_s},
\end{equation}
where $c$ is the speed of light. Therefore, the velocity variation
can be related to the matter-energy content of the universe by
setting $a(t_0)=1$ and using the Friedmann equation,
\begin{equation}\label{eq6}
\frac{\Delta v}{c}=H_0\Delta t_0\bigg[1-\frac{E(z_s)}{1+z_s}\bigg],
\end{equation}
where $H_0$ is the Hubble constant and $E(z)=H(z)/H_0$.

The feasibility of detecting a time evolution of the redshift was
once studied in detail by Pasquini {\it et
al.}~\cite{Pasquini1,Pasquini2}. The most promising system used to
measure the velocity shift within the redshift desert is quasar
absorption lines typical of the Lyman-$\alpha$ forest. The European
Extremely Large Telescope with a high-resolution ultra-stable
spectrograph such as CODEX will be able to detect the tiny shift in
spectral line over a reasonable time interval, typically of the
order of few decades~\cite{EELT}.

According to the latest Monte Carlo simulations, the accuracy of the
spectroscopic velocity shift measurements expected by CODEX can be
expressed as
\begin{equation}\label{eq7}
\sigma_{\Delta_v}=1.35\frac{2370}{S/N}\sqrt{\frac{30}{N_{\mathrm{QSO}}}}
\bigg(\frac{5}{1+z_{\mathrm{QSO}}}\bigg)^x~\mathrm{cm}~\mathrm{s}^{-1},
\end{equation}
where $S/N$ is the signal-to-noise ratio, $N_{\mathrm{QSO}}$ is the
number of observed quasars, $z_{\mathrm{QSO}}$ represent their
redshift and the index $x$ is 1.7 for $z\leqslant4$ while it becomes
0.9 beyond that redshift. The mock SL data set used in our following
analysis corresponds to the error bars computed from Eq.~(\ref{eq7})
with a $S/N$ of 3000 and a number of QSO $N_{\mathrm{QSO}}=30$
assumed to be uniformly distributed among the following redshift
bins: $z_{\mathrm{QSO}}=[2.0,2.5,3.0,3.5,4.0,4.5,5.0]$. The fiducial
concordance cosmological model with the parameters taken to be the
best-fit ones from  WMAP nine years analysis~\cite{WMAP9} is applied
to examine the capacity of future measurements of SL test to
constrain the concerned models

\subsection{The Cosmic Microwave Background and Baryonic Acoustic Oscillations}
There are two useful parameters commonly employed when analyzing the
CMB observations. One describes the scaled distance to
recombination, $\mathcal{R}$, and the other the angular scale of the
sound horizon at recombination,
$\ell_A$~\cite{Komatsu,Elgaroy,Wang}.

The shift parameter $\mathcal{R}$ is defined as
\begin{equation}\label{eq8}
\mathcal{R}=\sqrt{\Omega_m}\int_0^{z_*}\frac{dz}{E(z)},
\end{equation}
where $z_*\sim1091$ is the redshift of the last-scattering surface.

The position of the first CMB power-spectrum peak, which corresponds
to the angular scale of the sound horizon at recombination, is given
by
\begin{equation}\label{eq9}
\ell_A=\pi\frac{d_A(z_*)}{r_s(z_*)},
\end{equation}
where $d_A$ is the comoving angular diameter distance, $r_s(z_*)$ is
the comoving sound horizon at recombination
\begin{equation}\label{eq10}
r_s(z_*)=\int_{z_*}^\infty\frac{c_s}{H(z)}dz
\end{equation}
which is dependent on the speed of sound, $c_s$, in the early
universe. Using both these two parameters in combination reproduces
closely the fit from the full CMB power spectrum and  it was shown
that constraints from the shift parameter $\mathcal{R}$ alone could
approximately represent the degeneracy directions between model
parameters from the full CMB observations~\cite{Elgaroy}.

In our analysis, the data based on measurements of the CMB acoustic
scale~\cite{WMAP9}
\begin{equation}\label{eq11}
\ell_A=302.35\pm0.65
\end{equation}
and the ratios of the sound horizon scale at the drag epoch
($z_d\sim1021$) to the BAO dilation scale~\cite{BAO2}
\begin{eqnarray}\label{eq12}
\begin{split}
\frac{r_s(z_d)}{D_V(z=0.20)}&=0.1905\pm0.0061&\\
\frac{r_s(z_d)}{D_V(z=0.35)}&=0.1097\pm0.0036&\\
\frac{r_s(z_d)}{D_V(z=0.44)}&=0.0916\pm0.0071&\\
\frac{r_s(z_d)}{D_V(z=0.60)}&=0.0726\pm0.0034&\\
\frac{r_s(z_d)}{D_V(z=0.73)}&=0.0592\pm0.0032&
\end{split}
\end{eqnarray}
where the so-called dilation scale, $D_V$, is given by
\begin{equation}\label{eq13}
D_V(z)=\bigg[(1+z)^2d_A^2\frac{cz}{H(z)}\bigg]^{1/3},
\end{equation}
are also taken into account to study the constraining power of SL
test.

By considering the ratio of the sound horizons at drag epoch and
photon decoupling, $r_s(z_d)/r_s(z_*)=1.045\pm0.012$~\cite{WMAP9},
and combining the observational results of (\ref{eq11}) and
(\ref{eq12}), we obtain
\begin{eqnarray}\label{eq14}
\begin{split}
f_{0.20}=\frac{d_A(z_*)}{D_V(z=0.20)}&=17.55\pm0.60&\\
f_{0.35}=\frac{d_A(z_*)}{D_V(z=0.35)}&=10.11\pm0.35&\\
f_{0.44}=\frac{d_A(z_*)}{D_V(z=0.44)}&=8.44\pm0.66&\\
f_{0.60}=\frac{d_A(z_*)}{D_V(z=0.60)}&=6.69\pm0.32&\\
f_{0.73}=\frac{d_A(z_*)}{D_V(z=0.73)}&=5.45\pm0.30.&
\end{split}
\end{eqnarray}
In addition, the coefficients $0.337$, $0.369$ and $0.438$ which
correlate the pairs of measurements at $z=(0.20,0.35)$,
$z=(0.44,0.60)$ and $z=(0.60,0.73)$ respectively are taken into
consideration in our analysis. It should be stressed that these
distance ratios is deemed to could provide more reasonable
constraints on non-standard dark energy models than the primitive
CMB or BAOs data~\cite{Sollerman,Zheng}.

\section{MODELS AND RESULTS}
In the last two decades or so, numerous models have been proposed to
explain the observed cosmic acceleration. Although some models may
be preferred with respect to others based on some statistical
assessment as they fit the data better with a small number of
parameters~\cite{Davis,Sollerman,Zheng}, most of them have not been
falsified by available tests of the background cosmology. These
models can be classified in two categories: (1)Models based on
an exotic component dubbed dark energy;
(2)Models based on modified gravity in which gravitational
interaction deviates from Einstein's theory of gravity. Examples of the latter  include DGP brane-world scenario~\cite{DGP}, $f(R$)
gravity~\cite{FR} and $f(T$) gravity~\cite{FT}. The SL test has been
applied to explore dark energy in the past few
years~\cite{Corasaniti,Balbi,Hongbao,Jinfei}. In this paper, we
focus on the time evolution of the cosmological redshift as a test of
modified gravity theories, including the predictions on the
time evolution of the velocity shift derived from observations
performed over a time interval $\Delta t_0=30~\mathrm{yr}$ and the
constraints on model parameters from a future
CODEX-like SL signal.

\subsection{DGP model}
The DGP model~\cite{DGP}, which provides a mechanism for accelerated
expansion without introducing a repulsive-gravity fluid, arises from
a class of brane-world scenario in which gravity leaks out into the
bulk above a certain cosmologically relevant physical scale. This
leaking of gravity is responsible for the increase in the expansion
rate with time. In the framework of a spatially flat DGP model, the
Friedmann equation is modified as
\begin{equation}\label{eq15}
H^2-\frac{H}{r_c}=\frac{8\pi G}{3}\rho_m,
\end{equation}
where $r_c=1/[H_0(1-\Omega_m)]$, which represents the critical
length scale beyond which gravity leaks out into the bulk. In this
paper, we investigate the generalized DGP model~\cite{DGP2}, which
interpolates between the pure $\Lambda$CDM and original DGP model
with an additional parameter $\lambda$,
\begin{equation}\label{eq16}
H^2-\frac{H^\lambda}{r_c^{2-\lambda}}=\frac{8\pi G}{3}\rho_m,
\end{equation}
where $r_c=1/[H_0(1-\Omega_m)^{\lambda-2}]$. Thus, we can directly
rewrite the above equation and obtain the expansion rate
\begin{equation}\label{eq17}
E^2(z)=\frac{H^2}{H_0^2}=\Omega_m(1+z)^3+(1-\Omega_m)E^\lambda(z).
\end{equation}
For $\lambda=1$, this agrees with the original DGP Friedmann-like
equation, while $\lambda=0$ leads to an expansion history identical
to that of standard $\Lambda$CDM cosmology. The cosmological
constraints on this generalized scenario from current observations
were presented in detail in Ref.~\cite{junqing}.

\begin{figure}[htbp]
\centering
\includegraphics[width=0.5\textwidth, height=0.3415\textwidth]{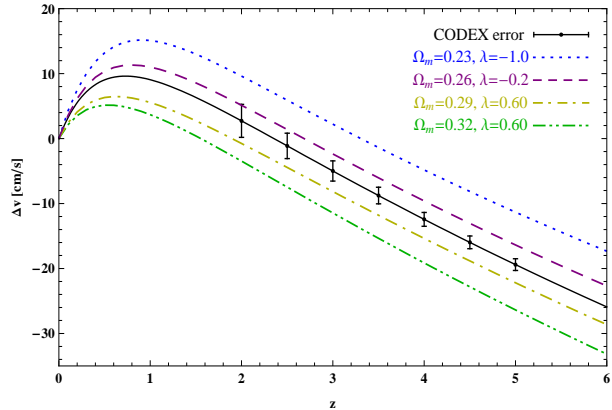}
\caption{\label{Fig1}
   The predicted velocity shift over a time interval $\Delta t_0=30~\mathrm{yr}$ for the DGP model,
   compared to simulated data as expected from the CODEX experiment. The mock data points and error bars are
   estimated from Eq.~(\ref{eq7}) with a fiducial concordance $\Lambda$CDM model.}
\end{figure}

\begin{figure}[htbp]
\centering
\includegraphics[width=0.45\linewidth]{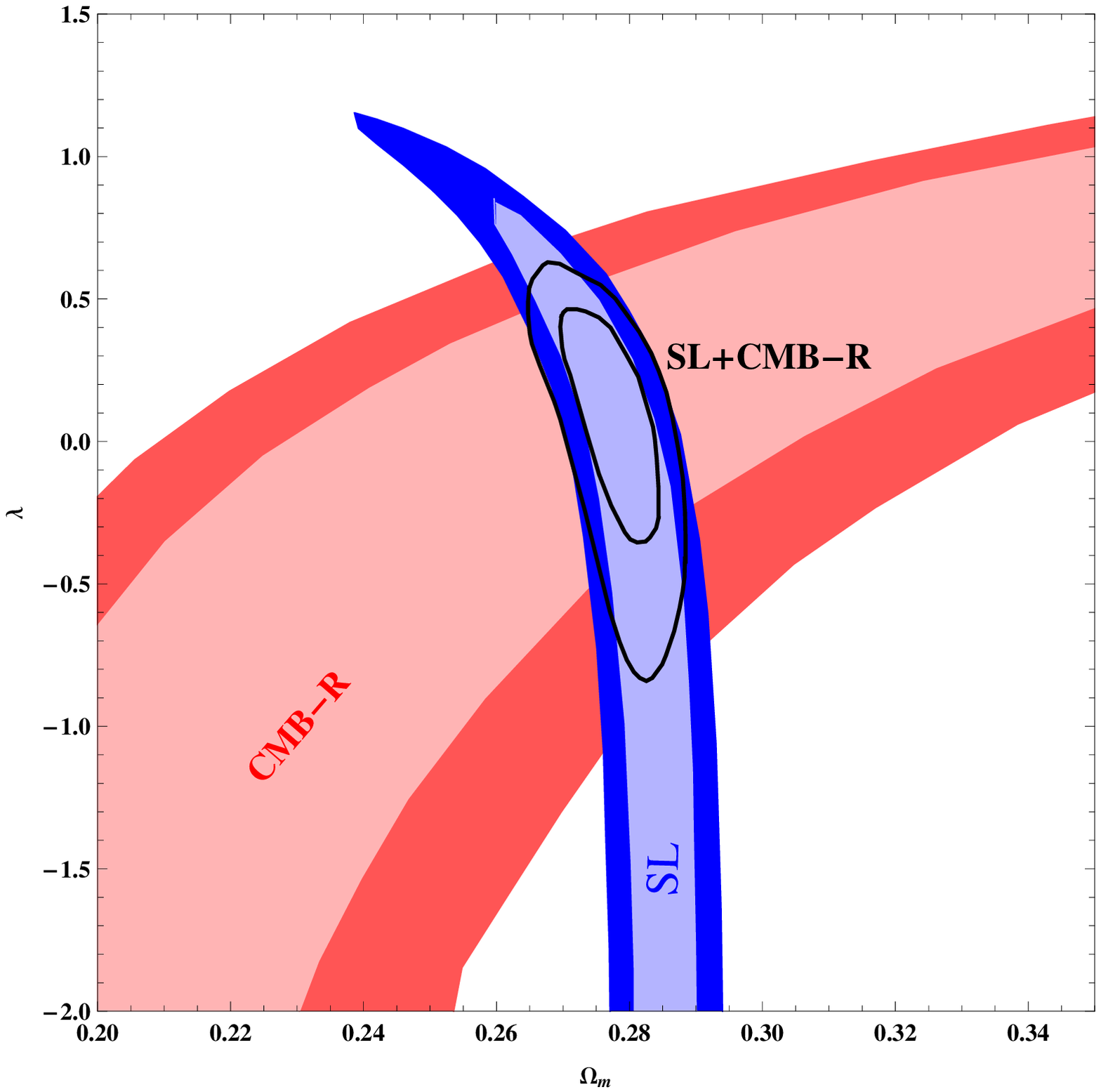}
\includegraphics[width=0.45\linewidth]{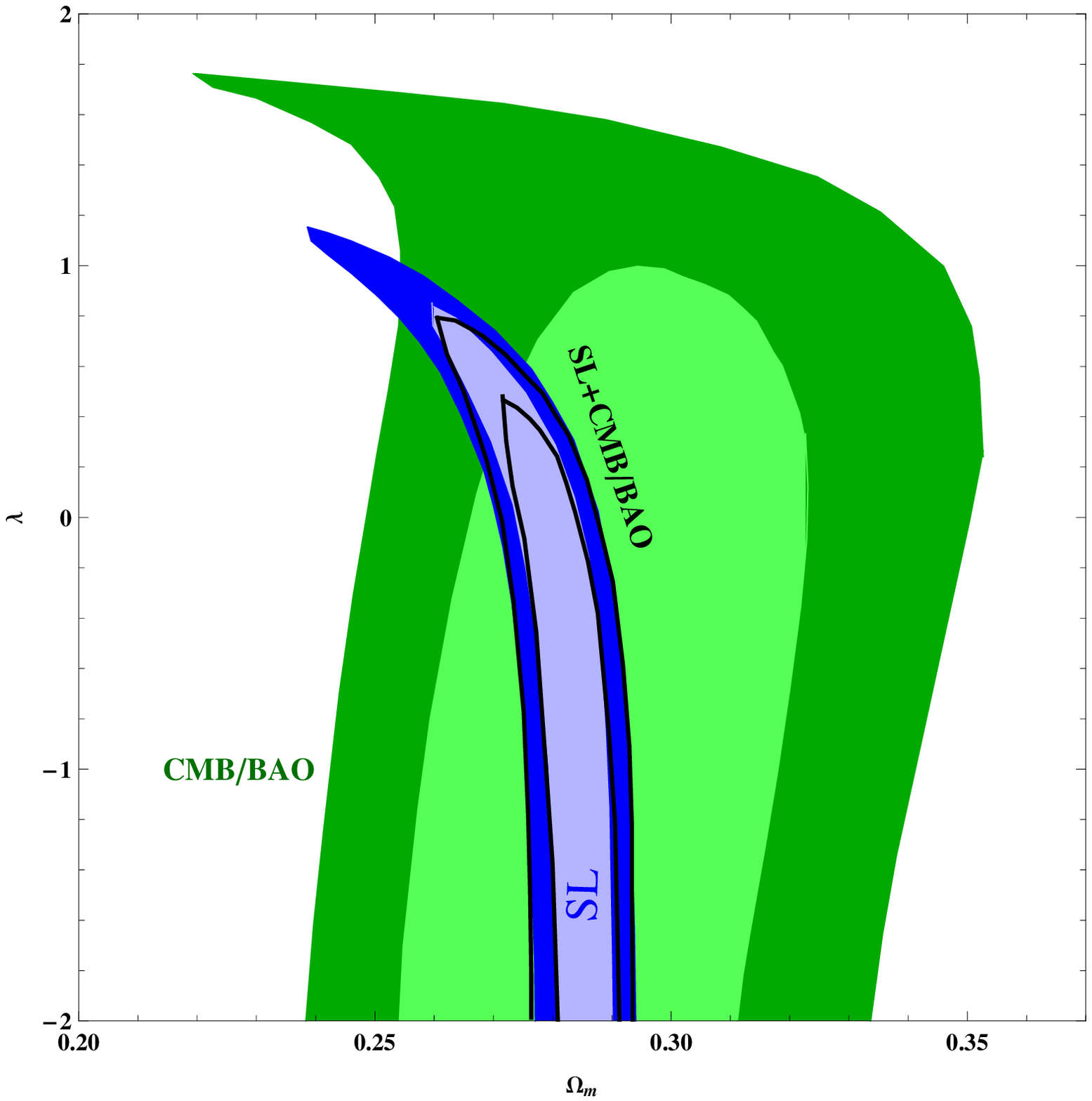}
\caption{\label{Fig2}
   $\mathrm{\mathbf{Left}}$: Constraints on the DGP model using SL test (blue layer), CMB--shift parameter (red layer), and combining the two probes (black solid lines).
   $\mathrm{\mathbf{Right}}$: Constraints on the DGP model using SL test (blue layer), CMB/BAO (green layer), and combining the two probes (black solid lines).}
\end{figure}

The predications on the time evolution of the velocity shift over a
time interval $\Delta t_0=30~\mathrm{yr}$ for the DGP model are
shown in Fig.~(\ref{Fig1}). Compared to simulated data as expected
from the CODEX experiment in a fiducial concordance $\Lambda$CDM
model with the parameters taken to be the best-fit ones from WMAP
nine years analysis~\cite{WMAP9}, the DGP model with
$\Omega_m\sim[0.23, 0.27]$ and $\lambda\sim[-0.2, 0.6]$ seems to be
favored. Alongside the primitive CMB--shift parameter or CMB/BAO,
the ability of a future SL signal measurement to constrain this
model is presented in Fig. (\ref{Fig2}). From the left panel of this
Figure, we find that the SL test is able to break the degeneracies
between model parameters when combined with the CMB--shift parameter
data and thus greatly improve cosmological constraints. This is
similar to what was obtained for a CPL-like dynamical dark
energy~\cite{Martinelli}. However, as shown in the right panel, the
advantage disappears when the CMB--shift parameter is replaced by
CMB/BAO. Nevertheless, in either case,  the SL test imposes a strong
bound on $\Omega_m$, which is similar to what was obtained when the
holographic dark energy model was
explored with the SL test~\cite{Hongbao}. 

\subsection{$f(R$) modified gravity}
The $f(R$) gravity theories modify general relativity by introducing
nonlinear generalizations to the (linear) Hilbert action (see
Refs.~\cite{FR1,FR2} for recent reviews). As the generalized
Lagrangians of this type can lead to accelerating phases both at
early~\cite{Starobinsky} and late~\cite{Capozziello1,Carroll} times
in the history of the universe (see also
Ref.~\cite{Capozziello3,Nojiri}), a great deal of interest and
effort has gone into the study of such theories.

The starting point of the $f(R$) theories in the Palatini approach
is the Einstein-Hilbert action, which is given by
\begin{equation}\label{eq18}
S=\int d^4x\sqrt{-g}\bigg[\frac{1}{2k}f(R)+\mathcal{L}_m\bigg],
\end{equation}
where $f$ is a differentiable function of the Ricci scalar $R$,
$\mathcal{L}_m$ is the Lagrangian of the pressureless matter,
$k=8\pi G$, and $G$ is the gravitational constant. For a flat
Friedmann-Lema\^{\i}tre-Robertson-Walker (FLRW) background, the
Hubble parameter in terms of the curvature scalar $R$ reads
\begin{equation}\label{eq19}
H^2=\frac{2k\rho_m+FR-f}{6F\xi},
\end{equation}
where
\begin{equation}\label{eq20}
\xi=\bigg[1-\frac{3F'(FR-2f)}{2F(F'R-F)}\bigg]^2,
\end{equation}
and $F=\partial f/\partial R$ and a prime denotes a derivative with
respect to $R$. In the case of the Hilbert action with $f=R$,
Eq.~(\ref{eq19}) reduces to the standard Friedmann equation:
$H^2=k\rho_m/3$. In this paper, we adopt the $f(R$) gravity theory
with the form $f(R)=R-\alpha(-R)^\beta$ within the Palatini approach
which can not only pass the solar system test and has the correct
Newtonian limit~\cite{Sotiriou}, but also can explain the late
accelerating phase of the expansion. Constraints from observations
such as the CMB--shift parameter, SNe Ia surveys data, BAOs, the
matter power spectrum from the SDSS and gravitational lensing on
this type of $f(R$) theory have been intensively
discussed~\cite{Santos,Fay,Borowiec,Amarzguioui,Sotiriou06a,Koivisto,LiB,DMchen,kai}.
Here, we evaluate the constraining power on this class of $f(R$)
gravity theory from a future measurement of SL test.

\begin{figure}[htbp]
\centering
\includegraphics[width=0.5\textwidth, height=0.3415\textwidth]{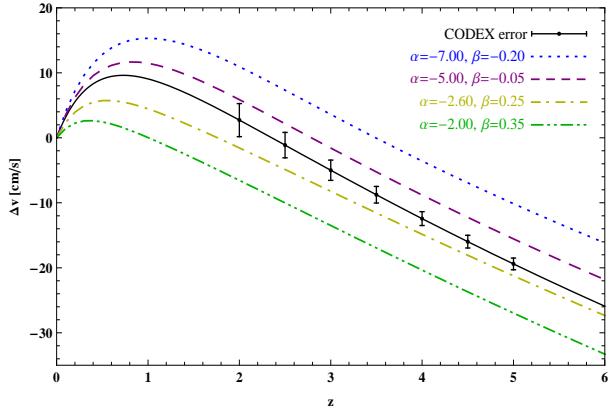}
\caption{\label{Fig3}
  The predicted velocity shift over a time interval $\Delta t_0=30~\mathrm{yr}$ for the model based on $f(R$) gravity,
   compared to simulated data as expected from the CODEX experiment. The mock data points and error bars are
   estimated from Eq.~(\ref{eq7}) with a fiducial concordance $\Lambda$CDM model. }
\end{figure}

\begin{figure}[htbp]
\centering
\includegraphics[width=0.45\textwidth, height=0.45\textwidth]{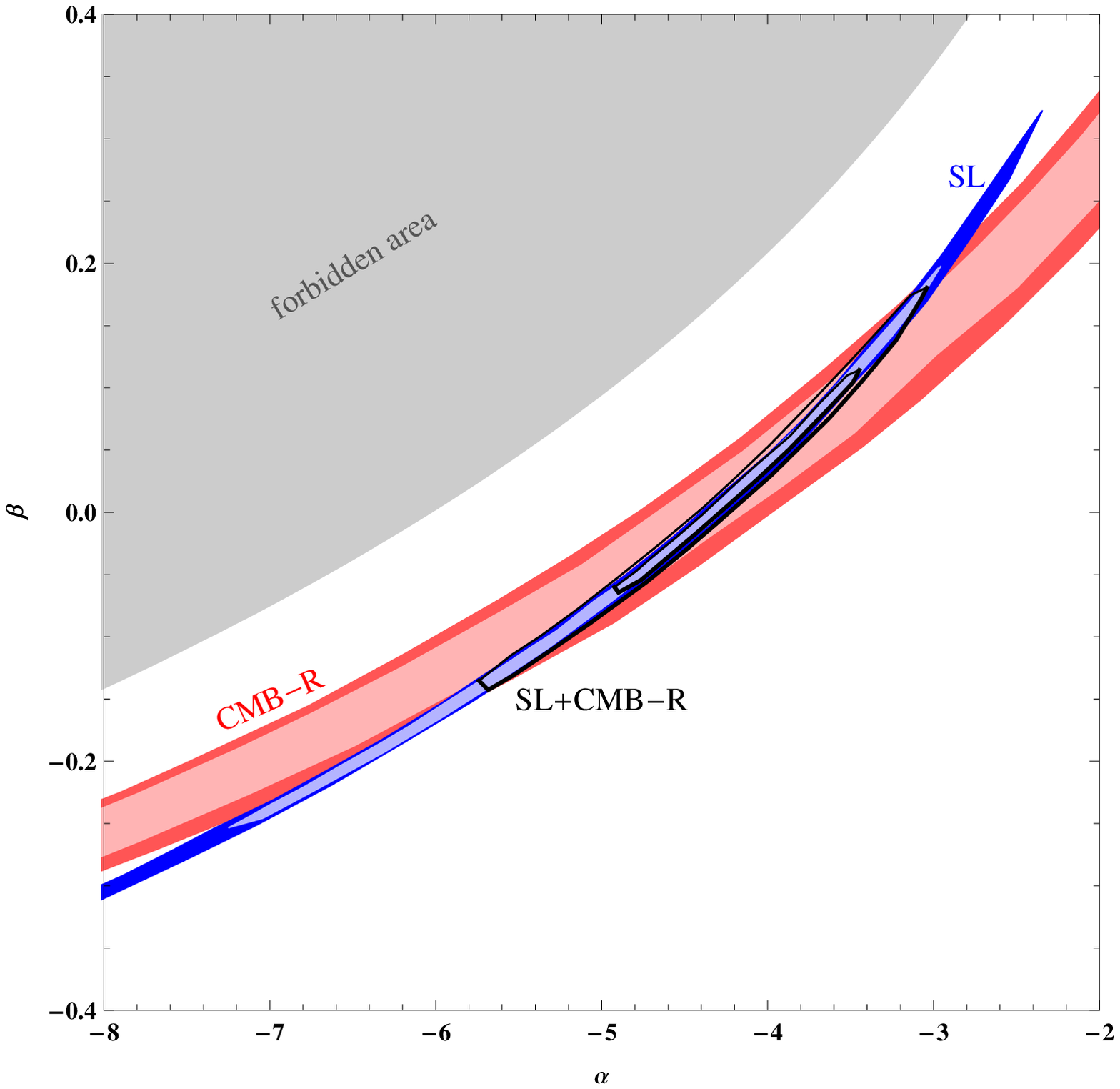}
\includegraphics[width=0.45\textwidth, height=0.45\textwidth]{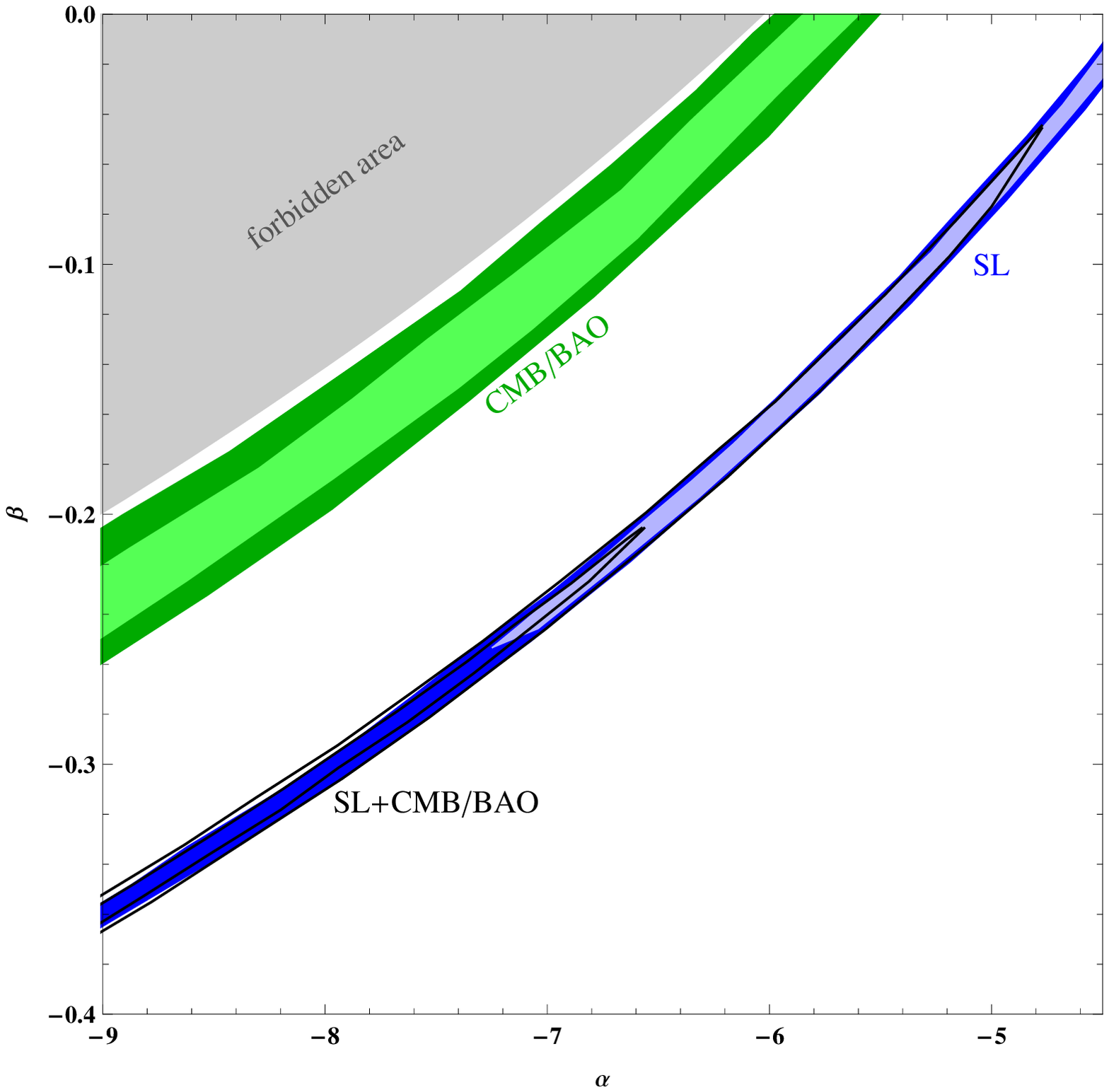}
\caption{\label{Fig4}
   $\mathrm{\mathbf{Left}}$: Constraints on the $f(R)$ gravity using SL test (blue layer), CMB--shift parameter (red layer), and combining the two probes (black solid lines).
   $\mathrm{\mathbf{Right}}$: Constraints on the $f(R)$ gravity using SL test (blue layer), CMB/BAO (green layer), and combining the two probes (black solid lines).}
\end{figure}

The predications on the time evolution of the velocity shift and the
numerical results which demonstrate the ability of a future SL
signal measurement to constrain this modified gravity theory are
shown in Fig.~(\ref{Fig3}) and Fig.~(\ref{Fig4}) respectively. The
results shown in the left panel of Fig.~(\ref{Fig4}) suggest that
the degeneracies between model parameters $\alpha$ and $\beta$ could
be broken by including the SL test when the observations of
CMB--shift parameter is considered. However, as shown in the right
panel, we find that the constraints from the SL test and the CMB/BAO
present almost the same directions of degeneracy between model
parameters. This means that, alongside the CMB/BAO, the SL test is
not capable of breaking the existing degeneracy between model
parameters.

It is interesting to note that the bounded regions in the
$\alpha-\beta$ plane from these two data sets are clearly different,
i.e., there is no overlap at 95.4\% confidence level (C. L.).
Therefore the SL test can provide complementary constraints on the
model parameters as compared to CMB/BAO, and the $\Lambda$CDM
($\beta=0$) might be ruled out at 95.4\% C. L. by the joint
analysis.

\subsection{$f(T$) gravity}
Recently, another kind of modified gravity, named $f(T)$ theory,
which can also explain the accelerating cosmic expansion, has
attracted an increasing deal of attention. In analogy to the $f(R)$
gravity, the $f(T)$ theory is obtained by extending the action of
teleparallel gravity which is based on teleparallel geometry where
the spacetime has only torsion and is
curvature-free~\cite{Einstein2,Hayashi1,Hayashi2}.

Assuming a flat FLRW metric, the expansion rate in terms of torsion
scalar $T$ is expressed as
\begin{equation}\label{eq21}
H^2=\frac{1}{6}\big[2k\rho-f+2Tf_T\big],
\end{equation}
where the subscript $T$ represents a derivative with respect to $T$
and $\rho$ is the energy density. 
 Recently, several specific
models based on $f(T)$ theory have been
proposed~\cite{Bengochea1,Linder}. Some of them can not only explain
the observed cosmic acceleration, but also can provide an
alternative to early inflation~\cite{Ferraro1,Ferraro2}.
Observational constraints and some important properties for these
models were extensively studied in the last few
years~\cite{Yang,Wu1,Wu2,Bengochea2,Chen,Rzheng,Bamba1}. In this
paper, we explore the $f(T)$ gravity theory with the SL test by
adopting a model, with the form
$f(T)=\alpha(-T)^n\tanh\frac{T_0}{T}$, in which the phantom divide
line crossing might be realized~\cite{Wu3}. The fundamental
requirement $\rho_{\mathrm{eff}}>0$ demands that the parameter $n$
must be greater than $3/2$.

\begin{figure}[htbp]
\centering
\includegraphics[width=0.5\textwidth, height=0.3415\textwidth]{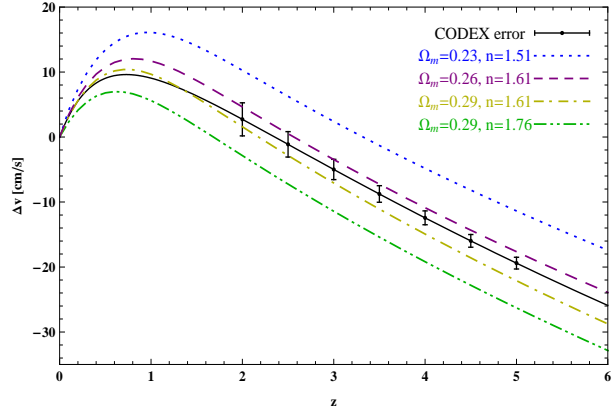}
\caption{\label{Fig5}
   The predicted velocity shift over a time interval $\Delta t_0=30~\mathrm{yr}$ for the model based on $f(T$) gravity,
   compared to simulated data as expected from the CODEX experiment. The mock data points and error bars are
   estimated from Eq.~(\ref{eq7}) with a fiducial concordance $\Lambda$CDM model.}
\end{figure}

\begin{figure}[htbp]
\centering
\includegraphics[width=0.45\textwidth, height=0.5\textwidth]{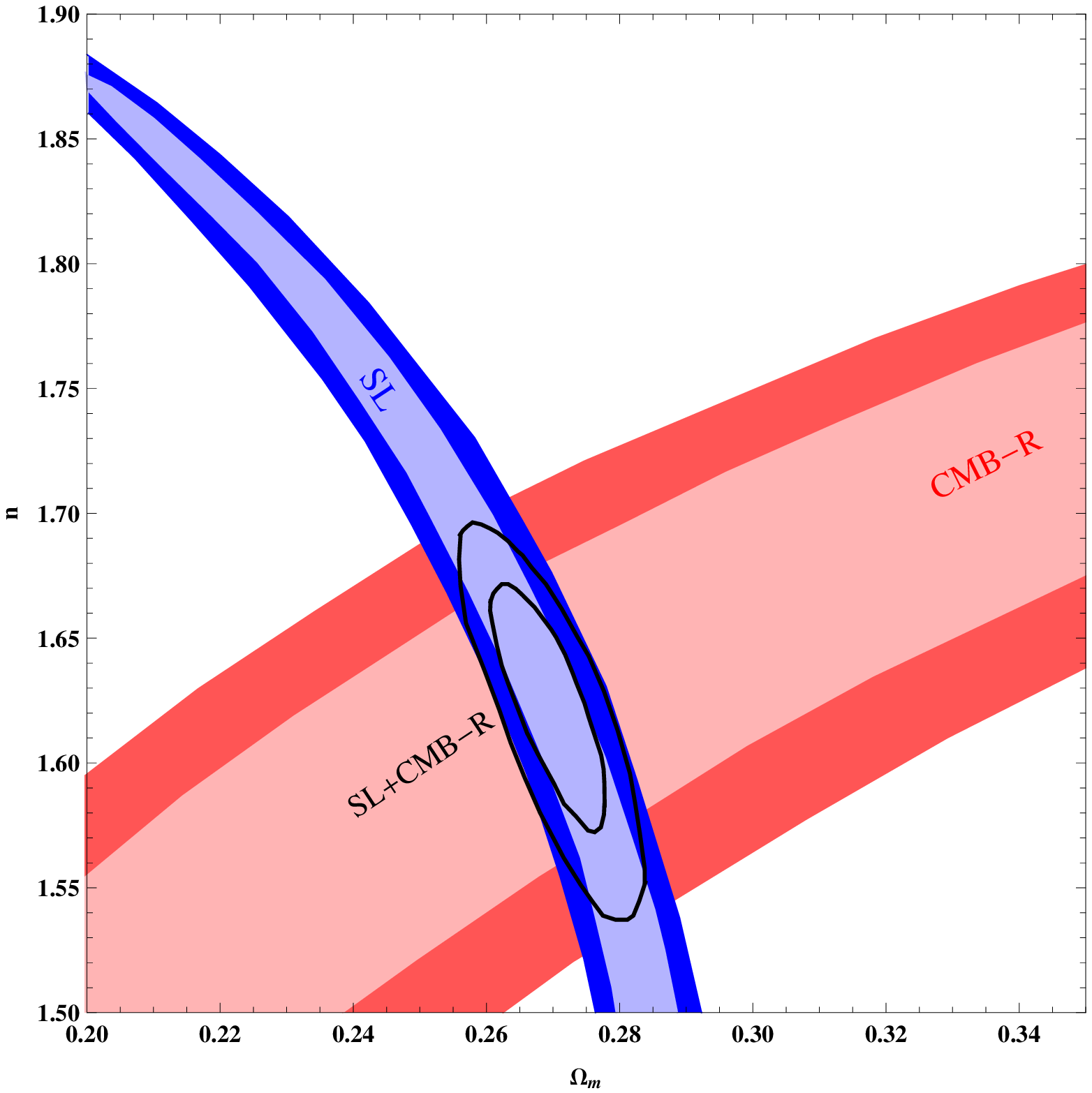}
\includegraphics[width=0.45\textwidth, height=0.5\textwidth]{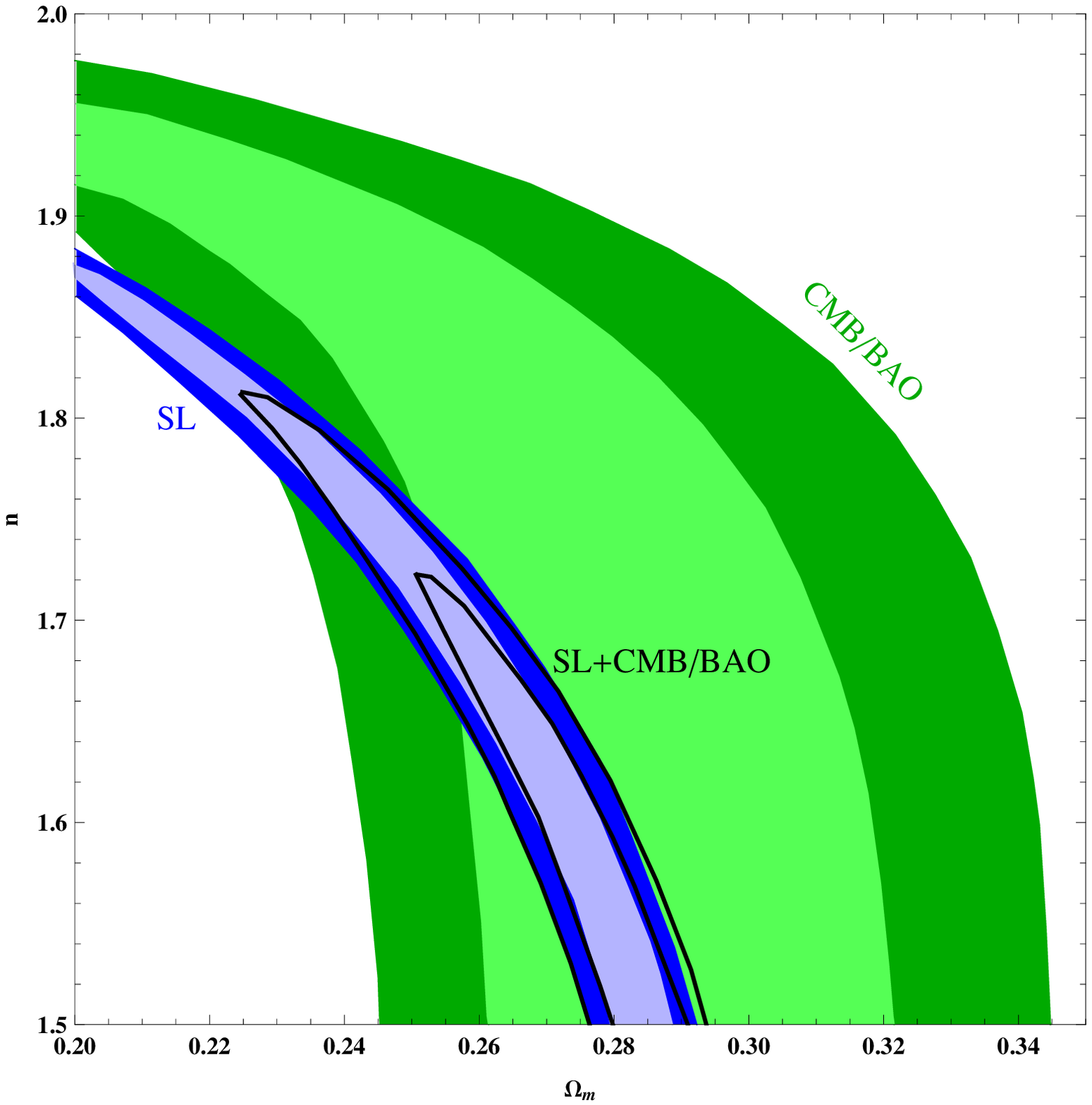}
\caption{\label{Fig6}
   $\mathrm{\mathbf{Left}}$: Constraints on the $f(T)$ gravity using SL test (blue layer), CMB--shift parameter (red layer), and combining the two probes (black solid lines).
   $\mathrm{\mathbf{Right}}$: Constraints on the $f(T)$ gravity using SL test (blue layer), CMB/BAO (green layer), and combining the two probes (black solid lines).}
\end{figure}

The predications on the time evolution of the velocity shift and the
numerical results which represent the ability of a future SL signal
measurement to constrain this modified gravity theory are shown in
Fig.~(\ref{Fig5}) and Fig.~(\ref{Fig6}) respectively.  The same as
for the two previously investigated modified gravity theories,
 the SL test  has a strong constraining power on the model parameters and greatly improve the
cosmological constraints when combined with the CMB--shift parameter data in
 the sense that it can
break the degeneracies between model parameters. However, this
constraining power disappears when the CMB--shift parameter is
replaced by the CMB/BAO.

\section{CONCLUSION}
In this paper, we have evaluated the  power of direct measurements
of temporal shift of cosmic redshift of the quasar spectra at
sufficiently separated epochs, i.e., the Sandage-Loeb (SL) test, on
constraining some popular modified gravity theories including DGP,
$f(R)$ modified gravity and $f(T)$ gravity theory. By considering
the signal from the Cosmic-Dynamic-Experiment-like spectrograph, we
quantify the ability of a future measurement of SL test to constrain
these modified gravity theories. Alongside the latest observations
of CMB--shift parameter, the SL test measurements are able to break
degeneracies between model parameters markedly and thus greatly
improve cosmological constraints for all investigated modified
gravity theories. This is similar to the constraining power of the
SL test on the phenomenological dynamic dark energy model with the
CPL parametrization~\cite{Martinelli}. In addition, the distance
ratios derived from the latest observations of the cosmic microwave
background and baryonic acoustic oscillations (CMB/BAO), which are
regarded to be more suitable than the primitive CMB--shift parameter
for testing non-standard dark energy models, are taken into
consideration for comparison. We find that the SL test measurements
and CMB/BAO yield almost the same directions of degeneracy between
model parameters for the concerned modified gravity theories. That
is, the inclusion of the SL test could not markedly improve the
constraints on these three modified gravity theories in terms of
degeneracy-breaking of model parameters when the CMB/BAO is
considered. However, for the DGP brane-world scenario and $f(T)$
gravity theory, due to a better sensitivity of the SL test, an
obvious improvement of constraint on the parameter $\Omega_m$ is
achieved, which is similar to what was obtained when the holographic
dark energy model was explored with the SL test~\cite{Hongbao}. This
advantage, of course, might result from the absence of systematic
effects which play a key role in the measurement of expansion
parameters. For the $f(R)$ modified gravity, the SL test can provide
completely different bounded regions in model parameters space as
compared to the CMB/BAO, and thus supplement strong complementary
constraints.

\section*{acknowledgments}
We would like to thank M. Martinelli for helpful discussions. This
work was supported by the Ministry of Science and Technology
National Basic Science Program (Project 973) under Grant
No.2012CB821804, the National Natural Science Foundation of China
under Grants Nos. 10935013, 11175093, 11075083 and 11222545,
Zhejiang Provincial Natural Science Foundation of China under Grants
Nos. Z6100077 and R6110518, the FANEDD under Grant No. 200922, the
National Basic Research Program of China under Grant No.
2010CB832803, the NCET under Grant No. 09-0144,  the PCSIRT under
Grant No. IRT0964, the Hunan Provincial Natural Science Foundation
of China under Grant No. 11JJ7001, and the SRFDP under Grant
No.20124306110001. ZL was partially supported by China Postdoc Grant
No .2013M530541.

\end{document}